# How Much Is Too Much? Measuring Divergence from Benford's Law with the Equivalent Contamination Proportion (ECP)

Manuel Cano-Rodríguez[*]

Department of Financial Economics and Accounting, University of Jaén, Jaén, Spain. ORCiD 0000-0002-3482-3836. E-mail mcano@ujaen.es

Abstract

Conformity with Benford's Law is widely used to detect irregularities in numerical datasets, particularly in accounting, finance, and economics. However, the statistical tools commonly used for this purpose—such as Chi-squared, MAD, or KS—suffer from three key limitations: sensitivity to sample size, lack of interpretability of their scale, and the absence of a common metric that allows for comparison across different statistics. This paper introduces the Equivalent Contamination Proportion (ECP) to address these issues. Defined as the proportion of contamination in a hypothetical Benford-conforming sample such that the expected value of the divergence statistic matches the one observed in the actual data, the ECP provides a continuous and interpretable measure of deviation (ranging from 0 to 1), is robust to sample size, and—under mild conditions—offers consistent results across different divergence statistics. Closed-form and simulation-based methods are developed for estimating the ECP, and, through a retrospective analysis of three influential studies, it is shown how the ECP can complement the information provided by traditional divergence statistics and enhance the interpretation of results.

[*]The author gratefully acknowledges the support by the Spanish Ministry of Science and Innovation under Grant PID2021-124494NB-I00.

# How Much Is Too Much? Measuring Divergence from Benford's Law with the Equivalent Contamination Proportion (ECP)

## 1. Introduction

Benford's law (BL) predicts the frequency distribution of the leading digits in numerical datasets arising from a wide range of natural and social phenomena. Its applicability has been documented across disciplines as diverse as physics and astrophysics, geology, medicine, economics, accounting and finance, and geography, among others. The widespread evidence of BL is evidenced by 907 citations of Benford's 1938 paper across 126 Web of Science categories as of the writing of this paper.

Business and economics category has become a prominent domain for BL applications, where it is commonly used to assess data quality and anomalies detection. The rationale is that naturally generated, unmanipulated data should conform to BL, while significant deviations may signal irregularities, noise or fraud. At the macroeconomic level, BL has been applied to assess the credibility of official statistics (Eutsler et al., 2023; Kaiser, 2019; Koch & Okamura, 2020; Michalski & Stoltz, 2013), and to detect illicit activities such as tax evasion, custom fraud, and money laundering (Ausloos et al., 2017; Badal-Valero et al., 2018; Barabesi et al., 2018; Nigrini, 1996). In accounting and finance, BL has been employed to flag potential errors or manipulations in financial reporting (Amiram et al., 2015; Beneish & Vorst, 2022; Chakrabarty et al., 2024), becoming a common technique in audit procedures and information systems for detecting anomalies (Cano-Rodríguez et al., 2025). More recently, it has also been applied to identify irregularities in cryptocurrencies and NFTs markets (Amiram et al., 2025; Cong et al., 2023; Sifat et al., 2024; Vičič & Tošić, 2022). In short, divergence from BL has become a widely adopted tool for evaluating data integrity in economic and financial contexts.

Studies assessing divergence from BL rely on a variety of statistical measures, such as the chi-squared test ($\chi^2$), the mean absolute difference (MAD), the sum of squared differences (SSD), the Euclidean distance (ED), the Kolmogorov–Smirnov test (KS), among others. These metrics are typically used either to test statistical significance—classifying datasets as conforming or not to BL—or to rank datasets by degree of conformity. However, this framework presents important limitations.

First, statistical significance is highly dependent on the dataset size $n$. In large datasets, even trivial deviations may produce statistically significant results (excess power problem), while in small datasets, substantial deviations may fail to reach significance (low power problem). Significance tests can therefore lead to misleading interpretations: large datasets with high conformity may be mistakenly classified as low-quality due to excessive test sensitivity (*fallacy of rejection*), while small datasets may appear reliable despite substantial deviations going undetected (*fallacy of acceptance*). To mitigate this, a more rigorous interpretation based on the severity principle is needed (Cerqueti & Lupi, 2023; Mayo & Spanos, 2006). This principle emphasizes that rejection is informative only if the test would not reject under near-conformity, and non-rejection is meaningful only if it would reject under relevant deviations.

Second, standard divergence metrics offer no absolute interpretation of how large the deviation from BL truly is. While they permit categorical or ordinal assessments, they lack a continuous, interpretable scale of magnitude. Some researchers (Kossovsky, 2021; Nigrini, 2012) have proposed cutoff points to classify conformity levels, but these thresholds are largely arbitrary and do not offer a continuous measure of the absolute magnitude of divergence.

Third, the wide variety of available metrics complicates comparison across studies. No clear consensus exists which statistic is superior, as each presents their own limitations and idiosyncracies (Campanelli, 2022, 2025; Cerasa, 2022; Cerqueti & Maggi, 2021; Kossovsky, 2021; Morrow, 2014). Furthermore, each operates on its own scale and responds differently to specific irregularities, complicating the comparison of results across studies.

To address these limitations, this paper proposes the Equivalent Contamination Proportion (ECP) as a complementary and informative measure of divergence from BL. The ECP quantifies the proportion of observations in a hypothetical BL-conforming dataset of the same size as the one under evaluation that would need to be replaced by non-BL values for the expected value of the divergence statistic to match the value observed in the actual dataset[1].

In this paper, estimation procedures for the ECP are developed for several commonly used divergence statistics using first-digit conformity and uniform contamination as the base case. However, the approach can be readily adapted to other contexts (conformity for second digit, last digit, first-two digits… or any other contamination distribution).

---

[1] For example, consider a dataset of 5,000 observations whose first-digit divergence from BL yields a chi-squared of 18 (significant at the 0.05 level). To compute the ECP, a hypothetical sample of the same size (5,000 observations) is created, in which a proportion of observations follows BL and the rest come from a non-BL distribution. The ECP is the proportion of non-BL values needed for the expected chi-squared statistic in the hypothetical sample to equal the observed value of 18.

The ECP has three key advantages. First, it is independent of the dataset size, mitigating both excess power and low power issues.[2] In severity-based terms, the ECP quantifies how much contamination would be needed to replicate the observed divergence—whether or not the null hypothesis is rejected—serving as a valuable input for avoiding both rejection and acceptance fallacies.

Second, the ECP is defined on a [0, 1] scale, where values near zero indicate minimal divergence and values near one suggest substantial deviation. This makes the ECP a continuous and interpretable measure of absolute divergence.

Third, it is easily computable from any divergence statistic, requiring only the observed statistic value, the dataset size, and the contamination distribution as inputs. Moreover, the results of this paper suggest that the ECP values derived from different statistics tend to be broadly congruent, although some variation remains depending on the nature of the deviations and each metric's sensitivity. This easy of computation and relative congruence across measures enable its retrospective application to existing studies, placing results on a common scale and facilitating comparison even when different divergence metrics are used.

It is worth emphasizing that the ECP is not a statistic in the conventional inferential sense: it does not follow a known sampling distribution under the null hypothesis and is not intended for hypothesis testing. Rather, it is an interpretable parameter that complements traditional tests by providing insight into the magnitude and practical relevance of observed deviations. The aim of the ECP, therefore, is to complement, not replace, standard hypothesis testing: while tests assess statistical significance, the ECP quantifies the substantive magnitude of the deviation.[3] To illustrate this complementary role in practice, this paper revisits three published studies that apply Benford's Law in different contexts. In each case, the retrospective estimation of the ECP provides additional insight into the magnitude and interpretability of the reported divergences, helping to refine or reassess the original conclusions.

---

[2] Although this independence from dataset size follows directly from the construction of the ECP—since it is defined on datasets of the same size—it is empirically confirmed in this paper through simulation in Section 3.1.

In sum, this paper introduces the ECP as a practical and complementary tool for assessing conformity with BL. By reframing divergence in terms of equivalent contamination, the ECP enhances interpretability, supports meaningful cross-study comparisons, and addresses critical limitations of current approaches.

The rest of the paper is structured as follows. Section 2 defines the ECP and describes how it can be estimated from different divergence statistics. Section 3 illustrates the properties and practical advantages of the ECP through simulation exercises and practical applications. Section 4 applies the ECP to three examples from prior research, highlighting how it can complement and refine the analysis of the divergence from BL. Section 5 concludes.

---

[3] For example, consider a dataset of 100 observations with a chi-squared statistic of 14 (below the threshold for significance at the 5% level). Under uniform contamination, the corresponding ECP is approximately 34.6%, indicating that a BL-conforming dataset of the same size with over a third of its values replaced by uniformly distributed ones would yield an expected chi-squared of 14. While the test suggests no statistical evidence against BL, the ECP shows that such a level of divergence could result from substantial contamination, thereby highlighting the risk of an acceptance fallacy. This type of insight is particularly valuable in severity testing (Mayo & Spanos, 2006), as it helps assess whether non-rejection reflects true conformity or a test's inability to detect meaningful deviations.

## 2. The Equivalent Contamination Proportion

The Equivalent Contamination Proportion (ECP) is defined as the proportion of contaminated observations in a hypothetical sample of random numbers of the same size as the evaluated dataset, such that the *expected* divergence from BL of the hypothetical sample equals the *observed* divergence of the empirical dataset, both measured using the same divergence statistic.

More formally, let $D$ be an empirical dataset with $n$ observations, and $T$ a divergence statistic (chi-squared, MAD, KS…). Denote $T(D)$ as the value of the statistic computed on $D$. Now consider a simulated sample $S$ of the same size $n$, in which a fraction of observations $f \in [0,1]$ follows a specified non-BL distribution (contaminated subsample), and the remaining $1-f$ follows the Benford distribution (non-contaminated subsample).

The ECP is then defined as:

$$ECP = f \in [0,1] : \mathbb{E}[T(S)] = T(D), \qquad (1)$$

where $\mathbb{E}[\cdot]$ denotes the expectation operator.

In other words, the ECP quantifies the contamination level that would be required in a perfectly BL-conforming sample for its expected divergence to match the value observed in the empirical dataset.

The procedures for deriving the Equivalent Contamination Proportion (ECP) for various commonly used divergence statistics—chi-squared, SSD, MAD, ED, KS, Kuiper, and CvM—are presented in detail in the online appendix. This includes both closed-form solutions (when available) and simulation-based estimation methods for non-additive statistics. The appendix also discusses implementation aspects and the rationale for adopting the uniform distribution as the contamination distribution.

In addition, the supplementary material includes a Python file (ECP.py) and an accompanying README file. This script can be used to estimate ECP values for leading-digit analysis using the aforementioned statistics, with the option to specify the uniform distribution (default) or any other user-defined alternative as the contaminant distribution.

## 3. Evaluating the properties of the ECP

This section evaluates the behavior of the ECP along two dimensions. First, I examine its robustness to sample size variations, specifically whether the ECP remains stable when the magnitude of the deviation from BL is held constant and the number of observations varies. If confirmed, this property would address one of the main methodological challenges in BL-based analysis. Second, I compare the ECP values from different divergence statistics applied to the same dataset. Since each statistic operates on a different scale and responds differently to data irregularities, cross-method comparisons are often problematic. If ECPs derived from different divergence statistics are congruent, this would support the use of the ECP as a unifying metric, translating heterogeneous outputs into a common, interpretable scale of equivalent data contamination.

### 3.1. Dataset size and ECP

To evaluate the robustness of the ECP to sample size, I generate random samples conforming to the BL and contaminate them with varying levels of contamination $f$. Specifically, I construct random samples of size 100, 1,000, 10,000, and 100,000, and apply contamination levels of 1%, 5%, 25%, 75%, 95%, and 99%. For each combination of sample size and contamination level, I simulate 100,000 random samples and compute the seven divergence statistics considered in this paper (Chi-squared, SSD, MAD, ED, KS, Kuiper, and CvM). I then compute the average value of each statistic across simulations and derive the corresponding ECP using that average and the corresponding sample size. The results of this procedure are presented in Table 1.

Table 1. Simulation results assessing the robustness of the ECP to sample size

|  |  | $n$ | 100 | 1000 | 10000 | 100000 | 100 | 1000 | 10000 | 100000 |
|---|---|---|---|---|---|---|---|---|---|---|
|  |  |  | \multicolumn{4}{c}{$f$=1%} | \multicolumn{4}{c}{$f$=5%} |
| Chi-squared | | Stat. | 8.0263 | 8.0586 | 8.4409 | 12.013 | 8.2509 | 9.1612 | 18.230† | 108.47* |
|  | | ECP | 0.68% | 0.84% | 1.00% | 1.00% | 4.61% | 4.95% | 5.00% | 5.00% |
| SSD | | Stat. | 0.0083 | 0.0008 | 0.0000 | 0.0000 | 0.0085 | 0.0009 | 0.0002† | 0.0001* |
|  | | ECP | 0.55% | 0.81% | 1.00% | 1.00% | 4.64% | 4.96% | 5.01% | 5.00% |
| MAD | | Stat. | 0.0235 | 0.0074 | 0.0024 | 0.0009 | 0.0237 | 0.0079 | 0.0037† | 0.0030* |
|  | | ECP | 1.03% | 0.59% | 0.99% | 0.99% | 4.52% | 4.86% | 5.00% | 5.00% |
| ED | | Stat. | 0.0878 | 0.0278 | 0.0090 | 0.0036 | 0.0886 | 0.0299 | 0.0143† | 0.0119* |
|  | | ECP1 | 0.00% | 0.00% | 0.00% | 0.90% | 0.00% | 3.29% | 4.76% | 4.97% |
|  | | ECP2 | 0.19% | 0.66% | 0.94% | 1.00% | 3.74% | 4.97% | 4.99% | 5.00% |
| KS | | Stat. | 0.0653 | 0.0207 | 0.0070 | 0.0035 | 0.0666 | 0.0247 | 0.0154* | 0.0136* |
|  | | ECP | 1.10% | 1.10% | 1.00% | 1.00% | 4.62% | 4.69% | 5.03% | 5.01% |
| Kuiper | | Stat. | 0.0821 | 0.0259 | 0.0085 | 0.0036 | 0.0831 | 0.0286 | 0.0155† | 0.0136* |
|  | | ECP | 2.38% | 0.93% | 1.06% | 1.00% | 7.16% | 4.80% | 4.90% | 5.01% |
| CvM | | Stat. | 0.0134 | 0.0013 | 0.0001 | 0.0000 | 0.0144 | 0.0022 | 0.0010* | 0.0009* |
|  | | ECP | 0.02% | 0.00% | 0.94% | 1.00% | 5.65% | 5.36% | 5.08% | 5.03% |
|  |  |  | \multicolumn{4}{c}{$f$=25%} | \multicolumn{4}{c}{$f$=75%} |
| Chi-squared | | Stat. | 11.309 | 33.930* | 259.92* | 2519.2* | 33.008* | 236.36* | 2269.3* | 22605* |
|  | | ECP | 24.66% | 24.97% | 25.00% | 25.00% | 74.88% | 74.99% | 74.99% | 75.00% |
| SSD | | Stat. | 0.0118 | 0.0042* | 0.0034* | 0.0034* | 0.0393* | 0.0314* | 0.0306* | 0.0305* |
|  | | ECP | 24.60% | 24.98% | 25.01% | 25.01% | 74.94% | 75.01% | 75.02% | 75.03% |
| MAD | | Stat. | 0.0278 | 0.0164* | 0.0150* | 0.0149* | 0.0501* | 0.0451* | 0.0447* | 0.0447* |
|  | | ECP | 24.11% | 24.96% | 25.00% | 25.00% | 74.52% | 75.00% | 74.99% | 75.00% |
| ED | | Stat. | 0.1049 | 0.0641* | 0.0589* | 0.0584* | 0.1955* | 0.1770* | 0.1750* | 0.1749* |
|  | | ECP1 | 21.22% | 24.46% | 24.96% | 25.00% | 73.54% | 74.86% | 75.00% | 75.03% |
|  | | ECP2 | 24.50% | 24.91% | 24.99% | 24.99% | 74.60% | 74.98% | 74.98% | 75.00% |
| KS | | Stat. | 0.0955 | 0.0719* | 0.0676* | 0.0672* | 0.2173* | 0.2029* | 0.2015* | 0.2015* |
|  | | ECP | 24.61% | 25.13% | 24.90% | 25.00% | 74.35% | 74.72% | 74.99% | 75.00% |
| Kuiper | | Stat. | 0.1038 | 0.0720* | 0.0676* | 0.0672* | 0.2177* | 0.2029* | 0.2015* | 0.2015* |
|  | | ECP | 25.51% | 25.20% | 25.01% | 25.00% | 74.43% | 74.74% | 74.99% | 75.00% |
| CvM | | Stat. | 0.0349 | 0.0226* | 0.0215* | 0.0212* | 0.2055* | 0.1924* | 0.1910* | 0.1909* |
|  | | ECP | 24.16% | 24.95% | 25.08% | 25.01% | 74.83% | 74.81% | 75.02% | 75.00% |

|            |       | f=95%   |         |         |         | f=99%   |         |         |         |
|------------|-------|---------|---------|---------|---------|---------|---------|---------|---------|
| Chi-squared | Stat. | 47.251* | 373.46* | 3636.1* | 36265*  | 50.565* | 404.71* | 3948.4* | 39381*  |
|            | ECP   | 94.91%  | 94.98%  | 95.00%  | 95.00%  | 99.01%  | 98.98%  | 99.00%  | 99.00%  |
| SSD        | Stat. | 0.0578* | 0.0499* | 0.0491* | 0.0491* | 0.0622* | 0.0541* | 0.0534* | 0.0533* |
|            | ECP   | 94.94%  | 95.03%  | 95.03%  | 95.04%  | 99.07%  | 99.02%  | 99.04%  | 99.04%  |
| MAD        | Stat. | 0.0608* | 0.0569* | 0.0567* | 0.0567* | 0.0631* | 0.0593* | 0.0591* | 0.0591* |
|            | ECP   | 94.50%  | 94.98%  | 95.00%  | 95.00%  | 98.62%  | 98.98%  | 99.01%  | 99.00%  |
| ED         | Stat. | 0.2382* | 0.2232* | 0.2216* | 0.2215* | 0.2472* | 0.2324* | 0.2309* | 0.2308* |
|            | ECP1  | 93.90%  | 94.92%  | 95.02%  | 95.04%  | 98.08%  | 98.92%  | 99.02%  | 99.04%  |
|            | ECP2  | 94.60%  | 94.91%  | 95.00%  | 95.00%  | 98.10%  | 98.93%  | 98.99%  | 99.00%  |
| KS         | Stat. | 0.2684* | 0.2560* | 0.2553* | 0.2553* | 0.2790* | 0.2667* | 0.2661* | 0.2660* |
|            | ECP   | 94.47%  | 94.70%  | 94.99%  | 95.00%  | 100.00% | 98.70%  | 99.00%  | 99.00%  |
| Kuiper     | Stat. | 0.2685* | 0.2560* | 0.2553* | 0.2553* | 0.2791* | 0.2667* | 0.2661* | 0.2660* |
|            | ECP   | 94.99%  | 95.10%  | 94.99%  | 95.00%  | 100.00% | 98.97%  | 99.00%  | 99.00%  |
| CvM        | Stat. | 0.3208* | 0.3077* | 0.3064* | 0.3063* | 0.3476* | 0.3340* | 0.3328* | 0.3327* |
|            | ECP   | 94.85%  | 94.73%  | 95.02%  | 95.00%  | 100.00% | 99.20%  | 99.02%  | 99.00%  |

Average values of the seven divergence statistics (Chi-squared, SSD, MAD, ED, KS, Kuiper, CvM) and correspondint estimated ECP for 100,000 simulated samples for each combination of sample size ($n$) and contamination level ($f$). For ED, two ECPs are reported: one from a closed-form approximation and one based on simulation. Simulation-based ECPs (ED, KS, Kuiper, CvM), use 5,000 random samples per iteration, with 95% confidence threshold guiding convergence. Symbols † and * denote values above the 95th and 99th percentiles of the null distribution, respectively, based on literature (Campanelli, 2025; Cano-Rodríguez et al., 2025; Morrow, 2014) or Monte Carlo simulation.

The results in Table 1 illustrate how traditional divergence statistics depend on sample size. At $n = 100$, none of the seven statistics is significant at the 5% level for $f \leq 25\%$. In contrast, even small deviations ($f = 5\%$) lead to significance when $n \geq 10{,}000$ observations. By comparison, the ECP is largely unaffected by sample size: for a given $f$, its estimated remains close to the true value across all statistics and $n$ values, confirming its robustness to sample size.

ECP estimates closely track the actual contamination levels introduced in the simulations. Minor deviations arise with small samples and low contamination, particularly for simulation-based ECPs, due to sampling variability and stochastic estimation. However, the errors are negligible in practical terms even for these cases. For instance, when n = 100 and $f$ = 1%, the estimated ECP is 2.38%, implying a relative error of 138% but only 1.38 percentage points in absolute terms.

Overall, Table 1 highlights the contrast between traditional statistics and ECP: while the former suffer from low power in small samples and excess power in large ones, the ECP remains stable and fairly accurate across all tested values of $f$ and $n$. This insensitivity to sample size reinforces its value as a scale-independent indicator of practical deviation from Benford's Law.

Beyond this general robustness, the availability of closed-form expressions for the Chi-squared, SSD, and MAD statistics allows computing, for each statistic, either (1) the minimum ECP required for the expected statistic to exceed the significance thresholds given a certain $n$, or (2) the minimum $n$ required for the expected statistic to exceed those thresholds for a given ECP. Table 2 summarizes these results: Panel A shows, by sample size, the minimum ECP required to exceed the 95% and 99% thresholds; Panel B shows, by ECP, the minimum $n$ required to exceed the same thresholds.

Table 2. Minimum ECP or Sample Size Required for the Expected Statistical to Reach Significance.

| | Chi-squared | | SSD | | MAD | |
|---|---|---|---|---|---|---|
| Panel A. ECP required for the expected statistic to achieve significance for the indicated sample size | | | | | | |
| **Sample size** | **p95** | **p99** | **p95** | **p99** | **p95** | **p99** |
| 100 | 39.14% | 50.78% | 38.85% | 51.10% | 42.95% | 55.06% |
| 500 | 18.47% | 23.67% | 17.55% | 23.01% | 19.67% | 25.06% |
| 1000 | 13.23% | 16.91% | 12.44% | 16.30% | 13.99% | 17.79% |
| 5000 | 6.02% | 7.67% | 5.59% | 7.31% | 6.30% | 8.00% |
| 10000 | 4.28% | 5.44% | 3.95% | 5.17% | 4.47% | 5.66% |
| 50000 | 1.92% | 2.44% | 1.77% | 2.32% | 2.00% | 2.54% |
| 100000 | 1.36% | 1.73% | 1.25% | 1.64% | 1.42% | 1.79% |
| 500000 | 0.61% | 0.78% | 0.56% | 0.73% | 0.63% | 0.80% |
| 1000000 | 0.43% | 0.55% | 0.40% | 0.52% | 0.45% | 0.57% |
| Panel B. Sample size required for the expected statistic to achieve significance for the indicated ECP | | | | | | |
| 1% | 185,982 | 300,072 | 156,984 | 268,586 | 300,072 | 156,984 |
| 5% | 7,296 | 11,860 | 6,249 | 10,713 | 11,860 | 6,249 |
| 10% | 1,780 | 2,921 | 1,553 | 2,669 | 2,921 | 1,553 |
| 25% | 264 | 447 | 245 | 424 | 447 | 245 |
| 50% | 58 | 103 | 60 | 105 | 103 | 60 |
| 75% | 22 | 43 | 27 | 47 | 43 | 27 |
| 90% | 14 | 28 | 19 | 32 | 28 | 19 |
| 95% | 12 | 25 | 17 | 29 | 25 | 17 |
| 99% | 11 | 23 | 16 | 27 | 23 | 16 |

Panel A reports the minimum ECP required for the expected values of the Chi-squared, SSD, and MAD to exceed the 95% and 99% critical thresholds for different sample sizes; Panel B reports the minimum sample size required for the expected values of the statistics to reach significance at 95% or 99% confidence levels for different ECP levels.

Table 2 illustrates the sample size problem. Regarding low power, it shows that small samples require substantial deviations to expect significant statistics. For example, with $n = 100$, no contamination level below 38% would be expected to trigger rejection for any of the three metrics. Conversely, detecting 25% contamination would require more than 244 observations. In terms of excess power, very large samples lead to rejection even under minimal deviations: At n = 1,000,000, significance would be expected even for contamination levels below 0.5%. Thus, no matter how small the deviation, one can always find a large enough $n$ to reject the null.

These results illustrate how the ECP enhances the interpretation of the statistical tests. While the statistic reflects significance, the ECP quantifies the practical magnitude of the deviation. In small samples, the ECP helps assess whether a non-significant divergence may still be relevant. For example, with $n = 200$ and MAD = 0.0212, the null is not rejected, but the ECP is around 22%, suggesting a meaningful deviation. Conversely, with $n = 100,000$ and MAD = 0.0011, the null is rejected, yet the ECP is only 1.42%, pointing to a minor divergence.

The use of the ECP alongside the divergence statistic offers a solution to the sample size problem, improving upon earlier alternatives. Subsampling or resampling techniques are among these alternatives (Bergh, 2015; Cerqueti & Lupi, 2021). Since large datasets tend to systematically reject the null, a potential mitigation strategy is to test conformity to BL on smaller random subsamples drawn from the original dataset. However, this approach reduces sensitivity at the cost of arbitrariness: as Table 2 shows, the choice of subsample size is determinant for the rejection or not of the null. This creates a risk that outcomes reflect the subsampling design rather than the underlying properties of the data. For example, a researcher seeking to demonstrate non-significant divergence could simply select a subsample small ehough to produce a non-significant statistic. In contrast, the ECP provides a size-independent interpretation that avoids such vulnerabilities.

Another strategy used to address excess power involves relying on fixed cutoffs or thresholds for statistics such as the MAD or SSD that offer a quick heuristic for assessing conformity. These thresholds aim to capture the practical relevance of deviations independently of formal significance tests, being the most popular those proposed by Nigrini (2012)[4] for the MAD and Kossovsky (2021) for the SSD[5].

---

[4] Nigrini's thresholds for the leading digit: MAD < 0.006, close conformity; 0.006 < MAD < 0.012, acceptable conformity; 0.012 < MAD < 0.015, marginally acceptable conformity; MAD > 0.015, non-conformity.

[5] Kossovsky's thresholds for the leading digit: SSD < 0.0002, Perfectly Benford; 0.0002 < SSD < 0.0025, Acceptable Close; 0.0025 < SSD < 0.0100, Marginally Benford; SSD > 0.0100, Non-Benford.

The use of these cutoffs faces two important limitations. First, they are heuristic and often calibrated for specific sample sizes and conditions, lacking solid theoretical grounding. Applying them across different sample sizes can lead to misclassifications (de Araújo Silva & Aparecida Gouvêa, 2023). Second, they rely on categorical scales (conformity, acceptable conformity, marginal conformity, non-conformity), which limits interpretive granularity.

Figure 1 illustrates how the SSD (panel A) and MAD (panel B) evolve with the ECP across sample sizes, alongside standard cutoff values. These plots reveal that the same threshold may correspond to very different contamination levels depending on $n$. For example, win $n \leq 1{,}000$, the expected MAD or SSD under BL (ECP=0) would never qualify as "close conformity" or "perfect Benford". Conversely, with $n \geq 5{,}000$, datasets with ECPs as high as 20–25% might still fall into "acceptable" ranges. Similarly, for SSD, ECP values below 39% would prevent a "non-Benford" label when $n \geq 500$. These patterns illustrate how thumb values may distort assessments across different settings.

Figure 1. Evolution of the SSD and MAD statistics as a function of the ECP

Panel A. SSD statistic

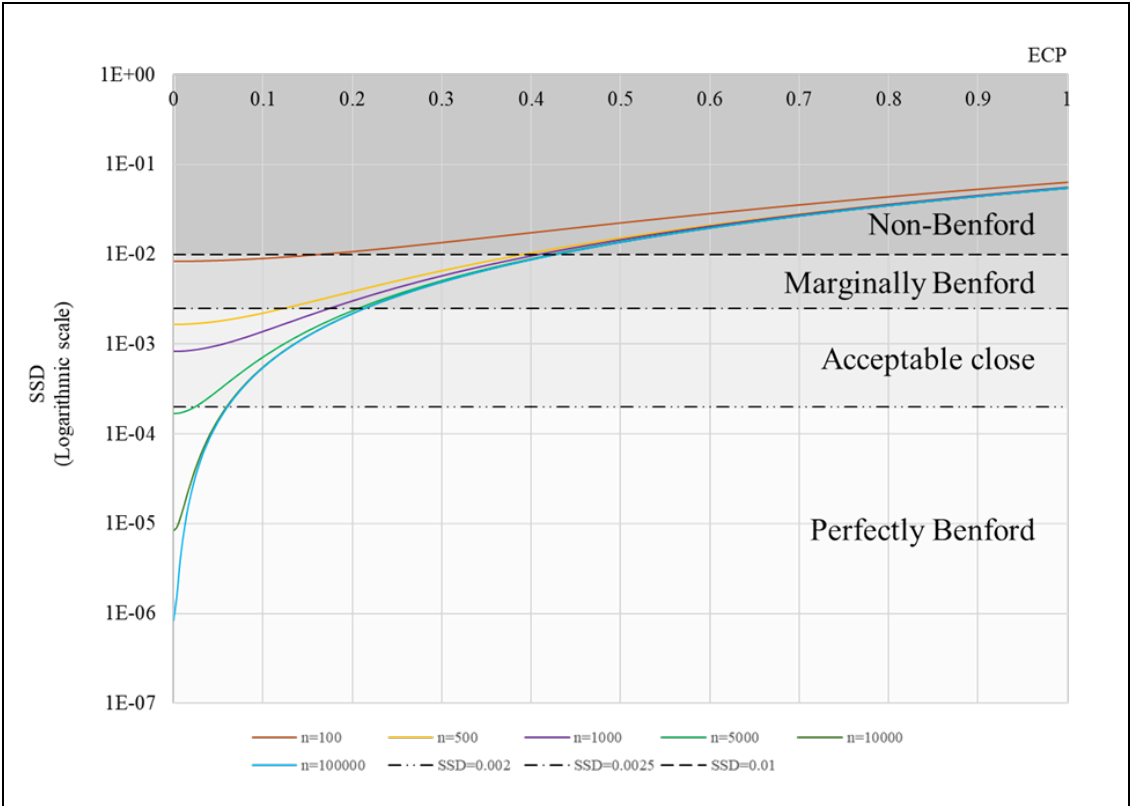

Panel B. MAD statistic

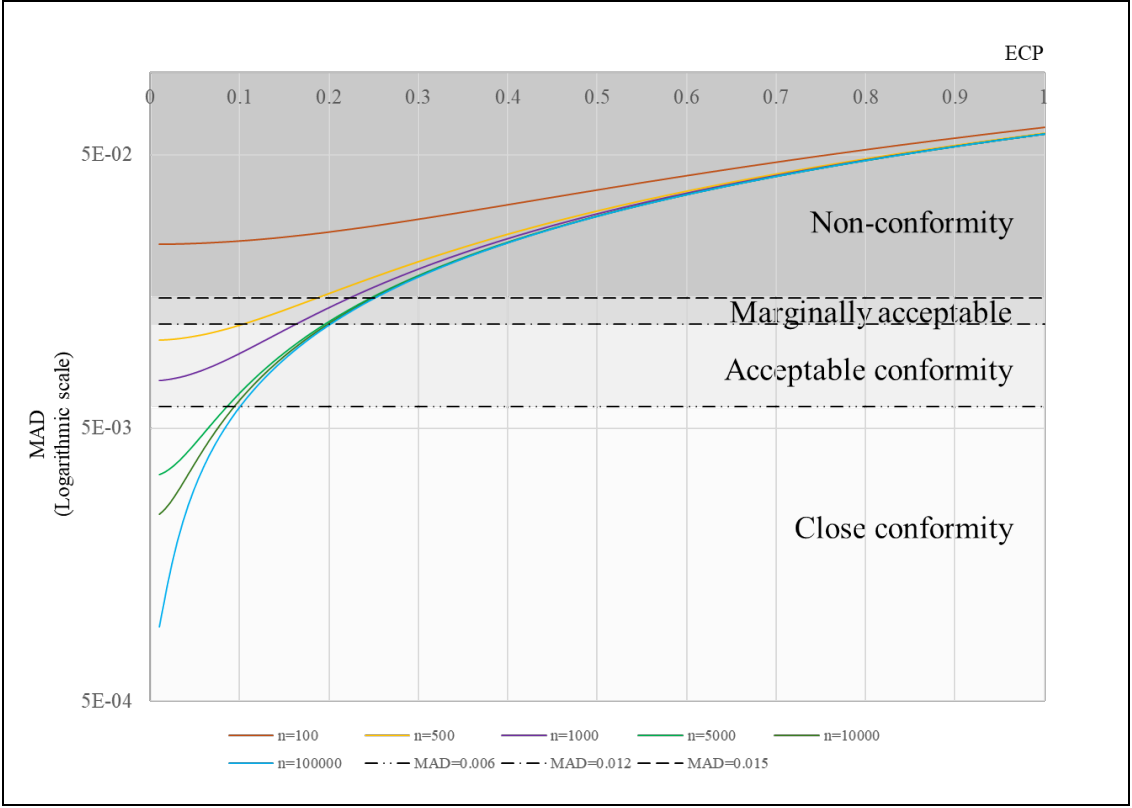

This figure shows how the SSD (panel A) and MAD (panel B) evolve with the ECP for sample sizes 100, 500, 50,00, 10,000, and 100,000. Horizontal axis indicates the ECP; vertical axis (log scale) display statistic values. Cutoffs from Kossovsky (2021) for the SSD and Nigrini (2012) for the MAD are also indicated.

Although both the ECP and cutoffs aim to capture the practical relevance of deviations, the ECP offers a continuous, scale-independent measure that avoids the arbitrariness and information loss of fixed thresholds, making it a more robust tool for conformity assessment.

In summary, the sample-size independence of the ECP makes it a valuable informational complement to traditional statistics, allowing consistent evaluation of BL deviations regardless of $n$. By mitigating both low and excess power issues, the ECP-statistic pairing helps prevent misleading conclusions. Compared to subsampling or heuristic cutoffs, the ECP provides a more consistent and theoretically grounded alternative.

### 3.2. Congruence of the ECP across statistics

In addition to its robustness to sample size, Table 1 reveals another key property of the ECP: the strong congruence across estimates derived from different divergence statistics. Despite differences in scale, sensitivity, and assumptions, the seven metrics yield nearly identical ECPs for the same datasets and contamination levels, highlighting the ECP's potential as a unifying, interpretable measure.

However, this congruence arises from simulated data in which the contamination pattern (uniform) exactly matches the model used for ECP estimation. In real-world datasets, contamination will be often unknown or more complex. To evaluate the congruence of ECP across statistics under such conditions, I apply the same seven statistics to 26 empirical datasets:

20 from Benford (1938) and 6 from Kossovsky (2021), the latter flagged as wrongly rejected as non-BL.[6]

Table 3 reports the divergence statistics for each dataset. These span in a wide range of sample sizes and conformity levels, making them suitable for testing ECP coherence. The overall level of agreement among the ECP estimates across the seven metrics is reasonably high. Table 4 confirms this with high pairwise correlations: ECPs from SSD, MAD, and ED are nearly perfectly aligned, and also highly correlated with those from chi-squared. ECPs from KS, Kuiper, and CvM are comparatively lower, yet consistently above 0.80, reinforcing the ECP's congruence across metrics.

---

[6] The original datasets used by Benford (1938) cover diverse domains, such as river surface areas, city populations, physical constants, molecular weights, and cost data. Divergence statistics were computing using first-digit frequences from his Table 1. The six datasets from Kossovsky (2021) include Time in Seconds between Global Earthquakes in 2012; USA Cities and Towns Population in 2009; Canford PLC price list from October 2013; Star Distances from the Solar System; DNA V230 Bone Marrow; and Oklahoma positive expenses below $1 million in 2011. They are available at S. J. Miller's Benford resources page:
https://web.williams.edu/Mathematics/sjmiller/public_html/benfordresources/

Table 3. ECP estimates for different empirical datasets

| Empirical dataset | n | Chi-squared | | SSD | | MAD | | ED | | KS | | Kuiper | | CvM | |
|---|---|---|---|---|---|---|---|---|---|---|---|---|---|---|---|
| | | Stat. | ECP | Stat. | ECP | Stat. | ECP | Stat. | Stat. | ECP | Stat. | ECP | Stat. | ECP | Stat. |
| Rivers, Area | 335 | 4.962 | 0.00% | 0.00134 | 0.00% | 0.0110 | 0.00% | 0.0366 | 0.00% | 0.0211 | 0.00% | 0.0300 | 0.00% | 0.0008 | 0.00% |
| Population | 3259 | 118.629 | 28.94% | 0.00389 | 25.86% | 0.0184 | 30.33% | 0.0624 | 26.02% | 0.0829 | 30.35% | 0.0829 | 30.38% | 0.0258 | 27.28% |
| Constants | 104 | 24.441 | 58.82% | 0.02683 | 58.17% | 0.0442 | 63.33% | 0.1638 | 59.61% | 0.1120 | 32.66% | 0.1722 | 56.73% | 0.0245 | 17.47% |
| Newspapers | 100 | 0.160 | 0.00% | 0.00012 | 0.00% | 0.0031 | 0.00% | 0.0111 | 0.00% | 0.0051 | 0.00% | 0.0080 | 0.00% | 0.0001 | 0.00% |
| Spec. Heat | 1389 | 111.213 | 42.70% | 0.00967 | 40.81% | 0.0269 | 44.09% | 0.0983 | 41.05% | 0.0610 | 21.25% | 0.1209 | 44.22% | 0.0127 | 18.57% |
| Pressure | 703 | 1.270 | 0.00% | 0.00016 | 0.00% | 0.0036 | 0.00% | 0.0127 | 0.00% | 0.0098 | 0.00% | 0.0149 | 0.00% | 0.0003 | 0.00% |
| H.P. Lost | 690 | 3.461 | 0.00% | 0.00038 | 0.00% | 0.0053 | 0.00% | 0.0195 | 0.00% | 0.0169 | 0.00% | 0.0179 | 0.00% | 0.0009 | 0.00% |
| Mol. Wgt. | 1800 | 125.757 | 40.12% | 0.00930 | 40.31% | 0.0258 | 42.51% | 0.0965 | 40.44% | 0.0820 | 29.53% | 0.1161 | 42.54% | 0.0240 | 26.09% |
| Drainage | 159 | 11.142 | 19.58% | 0.00764 | 20.41% | 0.0240 | 25.13% | 0.0874 | 22.64% | 0.0778 | 20.70% | 0.1079 | 32.70% | 0.0230 | 20.47% |
| Atomic Wgt. | 91 | 17.246 | 45.82% | 0.03850 | 72.80% | 0.0425 | 57.76% | 0.1962 | 73.75% | 0.1819 | 59.37% | 0.1911 | 63.43% | 0.0810 | 42.89% |
| $n^{-1}, \sqrt{n}, ...$ | 5000 | 440.764 | 46.33% | 0.00735 | 36.36% | 0.0253 | 42.30% | 0.0857 | 36.45% | 0.0872 | 32.14% | 0.0872 | 32.16% | 0.0318 | 30.50% |
| Design | 560 | 19.213 | 21.56% | 0.00372 | 20.10% | 0.0185 | 26.46% | 0.0610 | 21.09% | 0.0650 | 21.03% | 0.0650 | 20.73% | 0.0177 | 21.08% |
| Digest | 308 | 3.227 | 0.00% | 0.00175 | 0.00% | 0.0093 | 0.00% | 0.0418 | 0.00% | 0.0419 | 6.80% | 0.0419 | 7.01% | 0.0051 | 4.55% |
| Cost Data | 741 | 15.601 | 15.39% | 0.00211 | 13.33% | 0.0137 | 18.15% | 0.0459 | 14.29% | 0.0349 | 9.03% | 0.0349 | 7.19% | 0.0041 | 7.93% |
| X-Ray Volts | 707 | 5.860 | 0.00% | 0.00108 | 0.00% | 0.0086 | 0.00% | 0.0329 | 0.00% | 0.0231 | 0.00% | 0.0210 | 0.00% | 0.0014 | 0.00% |
| Am. League | 1458 | 14.595 | 10.31% | 0.00106 | 9.46% | 0.0073 | 6.85% | 0.0326 | 10.19% | 0.0280 | 7.81% | 0.0280 | 7.19% | 0.0041 | 9.64% |
| Black Body | 1165 | 43.032 | 26.99% | 0.00434 | 25.77% | 0.0132 | 19.18% | 0.0659 | 26.26% | 0.0679 | 23.58% | 0.0621 | 21.29% | 0.0177 | 22.06% |
| Addresses | 342 | 10.404 | 11.99% | 0.00177 | 0.00% | 0.0107 | 0.00% | 0.0421 | 0.00% | 0.0418 | 7.98% | 0.0538 | 23.52% | 0.0031 | 0.00% |
| $n^1, n^2, ... n!$ | 900 | 24.994 | 21.20% | 0.00366 | 22.32% | 0.0153 | 22.49% | 0.0605 | 22.89% | 0.0691 | 23.75% | 0.0691 | 23.52% | 0.0216 | 24.30% |
| Death Rate | 418 | 7.555 | 0.00% | 0.00248 | 9.20% | 0.0124 | 7.93% | 0.0498 | 10.98% | 0.0310 | 0.00% | 0.0420 | 4.96% | 0.0017 | 0.00% |
| Earthquakes | 19451 | 53.004 | 7.57% | 0.00031 | 7.05% | 0.0048 | 7.62% | 0.0177 | 7.05% | 0.0192 | 6.89% | 0.0216 | 7.62% | 0.0009 | 5.20% |
| USA population | 19509 | 17.524 | 3.46% | 0.00013 | 3.90% | 0.0031 | 4.51% | 0.0112 | 4.06% | 0.0087 | 2.60% | 0.0087 | 2.51% | 0.0003 | 2.37% |
| Canford prices | 15194 | 57.739 | 9.00% | 0.00053 | 9.37% | 0.0052 | 8.17% | 0.0231 | 9.37% | 0.0132 | 4.43% | 0.0180 | 6.26% | 0.0004 | 3.05% |
| NASA stars | 48111 | 538.107 | 16.55% | 0.00141 | 16.02% | 0.0109 | 18.20% | 0.0376 | 16.02% | 0.0490 | 18.22% | 0.0490 | 18.22% | 0.0079 | 15.24% |

| Dataset | N | | | | | | | | | | | | | |
|---|---|---|---|---|---|---|---|---|---|---|---|---|---|---|
| Bone marrow | 91223 | 34.686 | 2.69% | 0.00006 | 2.92% | 0.0019 | 2.90% | 0.0075 | 2.92% | 0.0045 | 1.39% | 0.0081 | 2.89% | 0.0001 | 1.18% |
| Oklahoma expense | 980190 | 978.424 | 4.96% | 0.00009 | 4.10% | 0.0025 | 4.25% | 0.0096 | 4.10% | 0.0068 | 2.51% | 0.0090 | 3.36% | 0.0001 | 1.86% |

This table reports seven divergence statistics (Chi-squared, SSD, MAD, ED, KS, Kuiper, and CvM) and their corresponding ECP estimates for 26 empirical datasets: the first 20 from Benford (1938) and the last 6 from Kossovsky (2021). The first column lists the dataset name, the second shows the sample size, and each pair of the subsequent columns gives the statistic and its ECP.

Table 4. Correlation coefficients for the ECP estimates across divergence metrics

| | ECP from Chi-squared | ECP from SSD | ECP from MAD | ECP from ED | ECP from KS | ECP from Kuiper | ECP from CvM |
|---|---|---|---|---|---|---|---|
| **ECP from Chi-squared** | | 0.9410 | 0.9721 | 0.9385 | 0.8909 | 0.9392 | 0.8433 |
| **ECP from SSD** | 0.9345 | | 0.9751 | 0.9996 | 0.9377 | 0.9461 | 0.8721 |
| **ECP from MAD** | 0.9240 | 0.9875 | | 0.9756 | 0.9064 | 0.9431 | 0.8571 |
| **ECP from ED** | 0.9219 | 0.9972 | 0.9854 | | 0.9359 | 0.9461 | 0.8708 |
| **ECP from KS** | 0.9665 | 0.9033 | 0.8870 | 0.8901 | | 0.9257 | 0.9650 |
| **ECP from Kuiper** | 0.9304 | 0.8812 | 0.8763 | 0.8760 | 0.9446 | | 0.8461 |
| **ECP from CvM** | 0.8892 | 0.9031 | 0.8815 | 0.8885 | 0.9338 | 0.8420 | |

Pairwise correlations across ECP estimates from the seven divergence metrics (based on the 26 datasets in Table 3). Pearson coefficients appear above the diagonal; Spearman coefficients below.

A closer look at Table 3 shows that in five datasets (Rivers, Area; Newspapers; Pressure; H.P. Lost; X-Ray Volts), all seven metrics unanimously indicate BL conformity, with ECPs equal to zero. In 12 additional datasets, the spread between the maximum and minimum ECP stays below 10%, suggesting strong agreement. Moderate coherence is observed in five datasets (ECP spread between 10% and 20%), while four datasets (Constants, Spec. Heat, Atomic Wgt., and Addresses) show greater divergence across the ECP values, likely due to differences in how each statistic captures deviations: SSD, MAD, and ED are more sensitive to average digit-level discrepancies; Chi-squared also capture pointwise differences, but gives more weight to higher digits. KS and CvM focus on cumulative deviations, making them less responsive to localized anomalies. Kuiper behaves as a hybrid case: cumulative yet more sensitive to individual-digit discrepancies. These differences explain occasional ECP variability despite general congruence.[7]

Overall, the ECP offers a reasonably coherent comparative scale across divergence measures. SSD, MAD, ED and to a lesser extent Chi-squared and Kuiper, tend to align closely. KS and CvM yield highly similar ECP values to each other, though they may diverge somewhat from the other group. Nevertheless, comparisons across metrics should be approached with caution, as digit-level discrepancy patterns can affect congruence.

---

[7] The Addresses dataset illustrates how differing metric sensitivities lead to varying ECP estimates. Digit-level deviations are minor for digits one through eight but substantial for digit nine. This results in zero ECP values for SSD, MAD, and ED. In contrast, the cumulative impact raises the KS and CvM statistics, yielding small but non-zero ECPs. The sharp anomaly in digit nine affects the chi-squared and especially the Kuiper statistic more strongly, resulting in the highest ECPs among all metrics.

## 4. Enhancing the analysis of the divergence from BL using the ECP. Application to three examples from prior literature

In addition to its robustness to sample size and its usefulness—with some limitations—for comparing conformity across different divergence metrics, the ECP has another key advantage: it requires only three elements for its estimation—the observed divergence statistic, the sample size, and the assumed contaminating distribution. This simplicity facilitates its retrospective application to prior studies, provided they report the statistic and sample size. In such cases, the ECP can enrich the original findings, refine their interpretation, or even prompt critical reassessments. To illustrate this potential, I revisit three studies that applied BL to detect accounting anomalies (Amiram et al., 2015), suspicious trading activity (Sifat et al., 2024), and COVID-19 data irregularities (Eutsler et al., 2023).

### *4.1. Amiram, Bozanic, and Rouen (2015)*

Amiram, Bozanic, and Rouen (2015) pioneered the use of BL divergence to detect errors or manipulations in firm-year level accounting data. Since the publication of this seminal work, the use of divergence from BL as a proxy for poor financial information quality at the firm-year level has steadily grown in popularity across accounting researchers.[8]

Amiram et al. (2015) computed the MAD and the KS statistics for the Compustat financial statements between 2001 and 2011, finding that 86% of the firm-year observations in their sample exhibited a non-significant KS statistic at the 5% level and concluding that "a significant majority of firm-year empirical distributions conform to Benford's Law" (Amiram et al., 2015, p. 1554).

---

[8] As of this manuscript's date, Amiram et al. (2015) has been cited in 89 papers on the Web of Science website.

However, this conclusion may exemplify a case of the *fallacy of acceptance* (Mayo & Spanos, 2006): inferring true conformity from non-rejection, without accounting for the test's power. Given the small average sample size (125 line items), the likelihood of failing to detect meaningful deviations is non-negligible, raising concerns about the severity of the test. Here, the ECP proves useful by quantifying the divergence magnitude.

Table 5 replicates their results by group size and reports the corresponding ECP values. These range from 19.12% to 45.86%, with an average of 34.73%, indicating substantial deviations. Moreover, under their rejection threshold (KS > $1.36/\sqrt{n}$), a sample with 34.53% contamination would still be classified as conforming, suggesting that many firm-years labelled as BL-consistent may exhibit non-trivial divergence when assessed using the ECP.

| | Average number of line items | Average MAD | ECP |
|---|---|---|---|
| Table 5. ECP values computed for Amiram, Bozanic, and Rouen (2015) | | | |
| Top 1% of line items | 169 | 0.0231 | 19.12% |
| Top tercile | 144 | 0.0259 | 28.23% |
| Middle tercile | 124 | 0.0292 | 33.59% |
| Bottom tercile | 108 | 0.0335 | 41.16% |
| Botomm 1% of line items | 100 | 0.0362 | 45.86% |
| Overall | 125 | 0.0296 | 34.73% |
| | | KS | |
| Rejection criterion | 125 | 0.1216 | 38.29% |

Average MAD values at the firm-year level by group size, as reported by Amiram et al. (2015, Table 1, Panel C). The last column of panel A shows the corresponding ECP estimates based on the reported MAD and number of line items.

In summary, combining the ECP with traditional tests improves the assessment of conformity to BL. While Amiram et al. (2015) conclude that most firm-year statements conform to BL based on non-significant test results, the ECP suggests this conformity may be overstated. The relatively high ECP values indicate that non-rejection likely stems from limited statistical power rather than true adherence to BL. By quantifying the magnitude of the divergence, the ECP refines the interpretation, helping distinguishing between genuine conformity and undetected deviations.

### *4.2. Sifat, Tariq, and Van Donselaar (2024)*

Sifat, Tariq, and Van Donselaar (2024) use conformity with BL to detect suspicious trading activities in NFT markets, continuing the research line on the conformity of financial market data (Ley, 1996; Nigrini, 2015; Riccioni & Cerqueti, 2018), particulary on cryptocurrencies and digital asset markets (Amiram et al., 2025; Cong et al., 2023; Vičič & Tošić, 2022).

They find strong divergence form BL in both price and volume data across most NFT markets, suggesting widespread manipulation in these markets. Given the large datasets typical in financial markets, such studies often face the excess power problem. Researchers have attempted to circumvent this issue by employing subsampling techniques to artificially reduce the sample size (e.g., Cong et al., 2023) or using cutoffs. Sifat et al (2024) rely on cutoffs, providing a clear case to examine the limitations of the threshold-based assessment. Table 6 reports the minimum and maximum ECPs corresponding to each conformity category used in their analysis (Close, Acceptable, Marginally Acceptable, and Nonconformity), based on their first-digit results from Tables 12, 14, 16, and 18.

| Table 6. ECP values computed for Sifat, Tariq, and Van Donselaar (2024) | | | |
|---|---|---|---|
| Category | Number of datasets | Min ECP | Max ECP |
| Table 12. Price data | | | |
| Close | 1 | 4.92% | 4.92% |
| Acceptable | 10 | 0.00% | 18.90% |
| Marginally acceptable | 5 | 20.29% | 22.88% |
| Nonconformity | 34 | 25.14% | 100.00% |
| Table 14. Volume data | | | |
| Close | 0 | n/a | n/a |
| Acceptable | 8 | 4.92% | 16.2872% |
| Marginally acceptable | 5 | 0.00% | 23.4375% |
| Nonconformity | 37 | 20.29% | 100.0000% |
| Table 16. Aggregated price data | | | |
| Close | 1 | 8.21% | 8.21% |
| Acceptable | 3 | 14.90% | 18.76% |
| Marginally acceptable | 0 | n/a | n/a |
| Nonconformity | 1 | 26.29% | 26.29% |
| Table 18. Aggregated volume data | | | |
| Close | 0 | n/a | n/a |
| Acceptable | 0 | n/a | n/a |
| Marginally acceptable | 0 | n/a | n/a |
| Nonconformity | 6 | 8.59% | 100.00% |
| Overall data | | | |
| Close | 2 | 4.92% | 8.21% |
| Acceptable | 21 | 0.00% | 14.90% |
| Marginally acceptable | 10 | 0.00% | 23.44% |
| Nonconformity | 78 | 8.59% | 100.00% |

This table presents the number of datasets for each of Nigrini's categories included in Tables 12, 14, 16, and 18 of Sifat et al. (2024), along with the maximum and minimum ECP values computed for each dataset category.

The analysis of these results illustrates how fixed thresholds can yield inconsistent classifications when applied across datasets of varying sizes. For price data (Table 16), the cutoffs perform reasonably well, with Acceptable datasets showing ECPs below 19%, Marginally Acceptable between 20–22%, and Nonconforming above 25%. However, one dataset with an ECP of 0% is only classified as Acceptable rather than Close Conformity.

For volume data, the categorization proves less reliable: Acceptable datasets have ECPs from 4.9% to 16.2%, Marginally Acceptable from 0% to 23%, and Nonconforming from 20% to 100%, resulting in clear overlap. A striking inconsistency arises in the aggregated data (Tables 16 and 18), where a dataset with an ECP of 8.21% is labelled Close Conformity, while another with 8.59% is deemed Nonconforming, highlighting the arbitrariness of fixed thresholds.

Overall, ECPs range from 4.92% to 8.21% for Close, from 0 to 14.90% for Acceptable, from 0 to 23.44% for Marginal, and from 8.59% to 100% for Nonconforming, with substantial overlap. This underscores how exceeding or falling below a given threshold may not consistently reflect the actual magnitude of the divergence.

Sifat et al (2024) also note discrepancies between Chi-squared and MAD results. In particular, Table 16 shows four cases—CyberKongz, Knight Story, Pixls Official, and Wrapped Cryptopunks—where the Chi-squared is not significant while the MAD exceeds the threshold for nonconformity. ECP estimates clarify these inconsistencies: for CyberKongz, the ECP is approximately 16–17% (first value based on Chi-squared, second on MAD); for Knight Story, 21–27%; for Pixls Official, 27–33%; and for Wrapped Cryptopunks, 31–48%. These results suggest that for the first case, the relatively low ECP makes the non-significant Chi-squared result more accurate than the MAD classification. For the other three cases, the higher ECP values indicate more substantial divergences from Benford's Law, thus supporting the conclusion of nonconformity suggested by the MAD statistic. Thus, the ECP helps arbitrate between conflicting tests by quantifying divergence on a common scale, reinforcing its interpretative value beyond fixed thresholds.

## *4.3. Eutsler, Norris, y Wozniak (2023)*

Eutsler, Norris, y Wozniak (2023) contribute to the research using BL conformity to assess data reliability in the context of COVID-19 (Balashov et al., 2021; Campolieti, 2021; Diekmann, 2007; Koch & Okamura, 2020; Kolias, 2022, etc.). Their study analyses COVID-19 case and death reports by U.S. counties, using MAD and SSD statistics to measure divergence from BL. They find greater deviations in counties with stronger Republican affiliations, interpreting this as evidence that political partisanship may have influenced reporting accuracy.

My reassessment focuses on the death counts data. Table 7 reproduces the sample size, MAD, and SSD values from Panels A and C of Table 2 in Eutsler et al. (2023), alongside the ECP estimates derived from those values. This allows for examining how the ECP can provide additional insights into the magnitude of the observed divergences.

| Table 7. ECP values computed for Eutsler, Norris, y Wozniak (2023) | | | | | |
|---|---|---|---|---|---|
| | n | MAD | SSD | ECP_MAD | ECP_SSD |
| Total reported deaths | 69287 | 0.0706 | 0.1096401 | 100% | 100% |
| Total reported deaths by political group | | | | | |
| RC/RG | 27065 | 0.0876 | 0.177759 | 100% | 100% |
| RC/DG | 15022 | 0.0854 | 0.16502915 | 100% | 100% |
| DC/DG | 15057 | 0.0465 | 0.04126317 | 77.87% | 87.11% |
| DC/RG | 12143 | 0.0461 | 0.04014396 | 77.20% | 85.90% |

This table reproduces the MAD and the SSD values reported in Panels A and C of Table 2 of Eutsler et al. (2023), and the ECP estimates derived from those MAD and SSD values.

The results presented in Table 7 show that both MAD and SSD reject the null hypothesis of BL conformity across all cases. While divergence levels are indeed higher in Republican-leaning counties in line with Eutsler et al's conclusions, MAD and SSD alone offer limited insight into the substantive magnitude of these deviations. By estimating the ECP, it becomes possible to quantify and more directly compare the degree of divergence across groups.

ECP estimates confirm the ordinal patter but also reveal the extreme magnitude of the deviations: the datasets for total reported deaths and Republican-affiliated counties yield ECPs of 100%, while Democratic-leaning counties show ECPs between 77% and 85%, depending on the statistic used. If one assumes that unmanipulated COVID-19 death data follow BL, these values would imply near-total contamination, comparable to datasets where nearly all values are replaced by uniformly distributed numbers.

Although manipulation cannot be ruled out, such extreme contamination levels make this explanation less plausible. An alternative plausible explanation is that unmanipulated death count data does not naturally conform to BL. As prior research suggests (Fewster, 2009), datasets constrained to narrow ranges or lacking multiple orders of magnitude do not usually conform to BL. Given that reported COVID-19 deaths are generally small relative to population size—and that many counties are sparsely populated—the observed nonconformity is likely structural. This may also explain why divergence is greater in Republican-leaning counties, which tend to be less populous on average.

In summary, while Eutsler et al. (2023) correctly identify a greater divergence in Republican counties, the lack of an absolute interpretative scale limits their analysis. The ECP not only confirms the ordinal differences but also indicates that the observed deviations may reflect structural data characteristics rather than manipulation. This illustrates the ECP's added value as a scale-based measure that enhances the interpretation of BL conformity beyond what traditional statistics offer.

## 5. Conclusions

Conformity with Benford's Law is widely used across many disciplines—particularly in business and economics—to detect anomalies in numerical data. However, traditional statistics (such as Chi-squared, MAD, KS, and others) suffer from three key limitations: (1) high sensitivity to sample size, which compromises the severity of testing; (2) scale dependence, making it hard to assess substantial relevance; and (3) lack of comparability across metrics.

This paper introduces the Equivalent Contamination Proportion (ECP) to address these issues. The ECP is defined as the proportion of contaminated data in a hypothetical BL-conforming sample such that the expected value of the divergence statistic matches the one observed in the actual data. It provides a continuous measure of absolute deviation on a scale that is both easily interpretable (a proportion between 0 and 1) and shared across different statistics. As demonstrated in this paper, the ECP is robust to sample size and, in general, yields reasonably coherent values across different divergence statistics. These properties make the ECP especially useful for comparing results obtained with different metrics or under different sample conditions.

It must be highlighted, however, that the ECP is not a test statistic in the inferential sense, but an estimated parameter. Therefore, its aim is not to compete with traditional hypothesis tests, but to complement them by providing additional insight into the magnitude and relevance of observed deviations.

The ECP can be estimated with only three inputs: the observed statistic, the dataset size, and a chosen contamination distribution. The supplementary material for this paper describes the methods for computing the ECP, as well as a Python script for its practical computation. These methods are focused on the analysis of the first significant digit, but they can be readily adapted to study any intermediate or trailing digit, or any set of digits.

The paper also illustrates the retrospective use of the ECP in three empirical studies. These applications show how the ECP enhances interpretation by quantifying divergence magnitude, resolving inconsistencies across statistics, and questioning the suitability of BL in certain contexts.

Nonetheless, the ECP has its own limitations. First, its value depends on the assumed contamination model, which must be chosen carefully to ensure interpretability. While the uniform distribution has served as a neutral benchmark for this paper, alternative models may be more appropriate in specific settings. Second, ECP estimates may vary across statistics depending on the divergence pattern and contamination model. This underlines the need to select both components thoughtfully to ensure meaningful comparisons.

**Data availability statement**

The data used in this study were either obtained from publicly available datasets hosted on S. J. Miller's Benford resources webpage at Williams College (https://web.williams.edu/Mathematics/sjmiller/public_html/benfordresources/), or extracted directly from the published tables of the cited articles.

**Use of generative AI**

I used ChatGPT (OpenAI, GPT-4, 2025) to assist in the preparation of this manuscript. Specifically, the tool was used to refine the clarity of the language in some sections and to help debug and streamline the Python code provided as supplementary material. All outputs generated by the AI were carefully reviewed and validated by the author to ensure their accuracy and integrity.

**Appendix. Derivation of the ECP for common divergence statistics.**

This appendix provides the full derivations of the Equivalent Contamination Proportion (ECP) for several commonly used divergence statistics, as referenced in Section 2 of the main text.

The ECP is defined in the section 2 of the main manuscript as follows:

$$ECP = f \in [0,1] : \mathbb{E}[T(S)] = T(D), \qquad (1)$$

Let $NB_d$ denote the probability of observing the leading digit(s) $d^9$ in the contaminated subsample, and $B_d$ the corresponding probability under BL. The difference between these probabilities is denoted by $\delta_d = NB_d - B_d$.

The probability of observing the leading digit(s) $d$ in the full sample $S$ is therefore:

$$p_d = f \cdot NB_d + (1-f) \cdot B_d = B_d + f \cdot \delta_d. \qquad (2)$$

Let $s_i$ denote the $i$-th element of $S$, with $i=1, 2 \ldots n$. Define the indicator function $F_d(s_i)$, which takes the value of 1 if the leading digit(s) of $s_i$ are equal to $d$, and 0 otherwise. This function follows a Bernoulli distribution with expected value $p_d$. The observed proportion of numbers with leading digit(s) $d$ in $S$ is

$$O_d = \frac{1}{n}\sum_{i=1}^{n} F_d(s_i). \qquad (3)$$

---

[9] $d$ may refer to a single digit or a set of digits, depending on the structure of the test. For instance, $d \in \{1, 2\ldots 9\}$ when analyzing the leading digit; $d \in \{0, 1, 2\ldots 9\}$ when analyzing an intermediate or trailing digit; $d \in \{10, 11\ldots 99\}$ when the two leading digits are considered, and so on. In summary, this framework applies to any subset of digits for which a theoretical distribution of expected frequencies is defined.

Since $O_d$ is the average of $n$ independent Bernoulli variables, it can be approximated by a normal distribution with the following mean and variance:

$$\mathbb{E}[O_d] = p_d \qquad (4)$$

$$[O_d] = \frac{p_d \cdot (1 - p_d)}{n}. \qquad (5)$$

Once the distribution of the digit probabilities in the hypothetical sample $S$ is defined, the next step consists in determining the functional form of the expected value of the divergence statistic $T$ computed on $S$, denoted $\mathbb{E}[T(S)]$. This expected value depends on three elements: (1) the sample size $n$, (2) the contamination proportion $f$, and (3) the probability distribution of the contaminated subsample, $NB_d$. Formally:

$$\mathbb{E}[T(S)] = g(n, f, NB_d), \qquad (6)$$

where $g(.)$ represents the functional relationship between the expected value of the statistic, the sample size, the contamination proportion, and the non-BL distribution adopted.

Once the functional form of $\mathbb{E}[T(S)]$ is established for a specific divergence statistic, the ECP value for the dataset $D$ corresponds to the value of $f$ that satisfies:

$$\mathbb{E}[T(S)] = g(n, f, NB_d) = T(D), \qquad (7)$$

considering that $n$ is the sample size of both $D$ and $S$.

In summary, $f$ represents the proportion of contaminated data required in a hypothetical sample of size $n$ such that its expected divergence equals the one observed in the empirical dataset. In other words, the divergence observed in $D$ is interpreted as being equivalent to the expected divergence of a BL-conforming sample partially contaminated in proportion *f* with observations drawn from the specified non-BL distribution.

Two important considerations must be taken into account when computing *f* to ensure that the ECP is well-defined and meaningful. By definition, *f* is a proportion bounded between 0 and 1. At its minimum value (*f* = 0), there is no contamination in the hypothetical sample, and the expected divergence statistic corresponds to that of a perfectly BL-conforming sample. However, due to sampling variability, empirical datasets that closely conform to BL may exhibit an observed divergence statistic lower than the expected value of the statistic under perfect BL conformity[10]. In such cases, equation [7] may admit a negative solution for *f*, or no solution at all. To address this, the ECP is defined as zero whenever the observed divergence statistic is lower than its expected value under full perfect BL conformity.

At the upper bound (*f* = 1), all the observations in the hypothetical sample follow the alternative (non-BL) distribution, and $\mathbb{E}[\theta_S]$ reaches its maximum under the specified contamination distribution. If the contamination distribution is too similar to BL, this maximum value of $\mathbb{E}[T(S)]$ may fall short of the observed statistic in certain empirical datasets. In such cases, equation [7] would admit no valid solution for *f* in the [0, 1] interval. When this occurs, two alternative approaches can be considered: (1) directly assigning the maximum value *f* = 1, or (2) selecting a different non-BL distribution that departs further from BL. In the extreme case, the distribution that maximizes divergence from BL is the degenerate distribution that concentrates all probability mass on the digit(s) with the lowest expected frequency under BL.[11]

---

[10] For instance, when analysing the first digit, the expected value of the chi-squared statistic for a perfectly BL-conforming sample is equal to 8. Consequently, the minimum value of $\mathbb{E}[T(S)]$ is 8. However, due to random variability in finite samples, the chi-squared statistic observed for an empirical dataset *D* that closely conforms to BL may fall below this expected value.

In the following subsections, the described method is applied to derive the ECP for several commonly used divergence statistics. For simplicity, and without loss of generality, the analysis focuses on the first (leading) digit, and the uniform distribution is adopted as the contamination distribution.

The selection of the uniform distribution is justified for several reasons. First, it has been widely used in prior BL-conformity studies to simulate data manipulation (Cano-Rodríguez et al. 2025; Cerqueti and Lupi 2021; Mumic and Filzmoser 2021), presumably because it is considered a generic and neutral option in simulation settings, consistent with the absence of expectations about particular digits being more likely in manipulated data. Second, although it does not represent the worst-case scenario in terms of deviation from BL, the uniform distribution still departs substantially from it (Berger and Hill 2011), ensuring that the maximum value of $\mathbb{E}[T(S)]$ is high enough to accommodate the values observed in most empirical datasets. Third, BL is often explained by emphasizing that, contrary to intuition, the distribution of the leading digits is not uniform (Geyer and Martí 2012; Kössler et al. 2024; Sambridge et al. 2010; Shao and Ma 2010). This suggests that a widespread misconception persists: that leading digits in real-world data should be uniformly distributed. Consequently, data manipulators unaware of BL may introduce distortions in the data consistent with a uniform distribution (Barabesi et al. 2023).

---

[11] For example, when testing conformity with a single-digit position—such as the first (leading) digit or the second digit—the degenerate distribution that maximizes divergence from Benford's Law assigns all probability mass to the digit 9, which typically has the lowest expected frequency under BL. In the case of two-digit analysis, this distribution would concentrate all mass on 99; for three digits, on 999; and so on.

Two distinct approaches are employed to derive the ECP, depending on whether the expected value of the divergence statistic under contamination, $\mathbb{E}[T(S)]$, can be obtained analytically or not. In the first case, $\mathbb{E}[T(S)]$ has a closed-form expression, which allows its value to be computed directly for any given contamination proportion *f*. This applies to statistics such as MAD, SSD, χ², and, under certain approximations, ED. In the second case, where no analytical expression for $\mathbb{E}[T(S)]$ is available, a simulation-based iterative method is used to estimate the ECP. Although computationally more intensive, this approach enables the estimation of the ECP for statistics such as the KS, Kuiper, and CvM, and can be extended to any divergence measure that varies continuously and monotonically with *f*.

### *A.1. Computation of ECP from the χ²*

The expected value of the chi-squared statistic for testing the conformity of the hypothetical sample *S* with BL is given by:

$$\mathbb{E}[\chi_S^2] = n \cdot \sum_{d=1}^{9} \mathbb{E}\left[\frac{(O_d - B_d)^2}{B_d}\right]. \tag{8}$$

Each term in the sum can be decomposed into two components using the mean-variance decomposition:

$$\mathbb{E}[\chi_S^2] = n \cdot \sum_{d=1}^{9} \mathbb{E}\left[\frac{(O_d - B_d)^2}{B_d}\right]. \tag{9}$$

Substituting the expected value and variance of $O_d$ (from expressions (4) and (5)) into equation (9) and simplifying:

$$\mathbb{E}\left[\frac{(O_d-B_d)^2}{B_d}\right] = \frac{[p_d-B_d]^2}{B_d} + \frac{p_d \cdot (1-p_d)}{n \cdot B_d} \quad (10)$$

Using expression (2) to substitute the value of $p_d$:

$$\mathbb{E}\left[\frac{(O_d-B_d)^2}{B_d}\right] = \left(1-\frac{1}{n}\right) \cdot f^2 \cdot \frac{\delta_d^2}{B_d} + \frac{f \cdot \delta_d \cdot \frac{1-2 \cdot B_d}{B_d} + (1-B_d)}{n} \quad (11)$$

Finally, substituting equation (11) into equation (8), the expected chi-squared statistic becomes:

$$\mathbb{E}[\chi_S^2] = (n-1) \cdot f^2 \cdot \sum_{d=1}^{9} \frac{\delta_d^2}{B_d} + f \cdot \sum_{d=1}^{9} \delta_d \cdot \frac{1-2 \cdot B_d}{B_d} + k - 1. \quad (12)$$

This expression shows that the expected chi-squared value increases linearly with the sample size $n$ and quadratically with the contamination proportion $f$, thereby offering valuable insight into how the statistic responds to variations in both parameters. Without requiring simulations, it enables the analytical determination of the minimum sample size needed for a given contamination level to produce a statistically significant result. Conversely, it also allows for identifying the minimum contamination level required, given a sample size, to exceed a specified threshold.

When the contamination distribution is assumed to be uniform ($NB_d = 1/9$), the following numerical approximations hold for the case of first-digit analysis: $\sum_{d=1}^{9} \frac{\delta_d^2}{B_d} \approx 0.4017$ and $\sum_{d=1}^{9} \delta_d \cdot \frac{1-2 \cdot B_d}{B_d} \approx 3.6153$. Substituting into equation [12]:

$$\mathbb{E}[\chi_S^2] \approx (n-1) \cdot 0.4017 \cdot f^2 + 3.6153 \cdot f + 8 \quad (13)$$

This quadratic equation in $f$ allows the ECP of any given dataset to be computed directly as the value of $f$ for which the expectation of $\chi_S^2$ equals the observed value $\chi_D^2$, given the sample size $n$.

*A.2. Computation of ECP from the SSD*

The sum of square differences for sample $S$ is computed as

$$\mathbb{E}[SSD_S] = \sum_{d=1}^{9} \mathbb{E}[(O_d - B_d)^2] \quad (14)$$

Each squared difference term can be decomposed using the mean-variance identity:

$$\mathbb{E}[(O_d - B_d)^2] = [\mathbb{E}(O_d - E_d)]^2 + \text{Var}[O_d] = [p_d - B_d]^2 + \frac{p_d \cdot (1 - p_d)}{n} \quad (15)$$

Substituting the value of $p_d$ from equation (2) and simplifying:

$$\mathbb{E}[(O_d - B_d)^2] = f^2 \cdot \delta_d^2 + \frac{B_d \cdot (1 - B_d) + f \cdot \delta_d \cdot (1 - 2 \cdot B_d) - f^2 \cdot \delta_d^2}{n} \quad (16)$$

Replacing equation (16) into (14), the expected SSD becomes:

$$\mathbb{E}[SSD_S] = \frac{1}{n} \cdot [f^2 \cdot (n-1) \cdot \sum_{d=1}^{9} \delta_d^2 + f \cdot \sum_{d=1}^{9} \delta_d \cdot (1 - 2B_d) + \sum_{d=1}^{9} B_d \cdot (1 - B_d)] \quad (17)$$

This expression shows that the expected *SSD* of sample *S* is a quadratic function of the contamination proportion *f*. It also reveals that its dependence on the sample size *n* is relatively weak, particularly for larger values of *n*, where the SSD tends to stabilize and the first term dominates the expression. This behavior supports the view of those authors who argue that the SSD is less sensitive to sample size effects than other statistics such as the chi-squared statistic (Campanelli 2022). However, for small samples, the influence of *n* may still be relevant, potentially limiting direct comparisons across samples of different sizes.

Assuming the uniform distribution as the contaminating distribution ($NB_d = 1/9$), the following numerical approximations apply: $\sum_{d=1}^{9} \delta_d^2 \approx 0.0543$, $\sum_{d=1}^{9} \delta_d \cdot$ $(1 - 2B_d) \approx 0.1087$, and $\sum_{d=1}^{9} B_d \cdot (1 - B_d) \approx 0.8345$. Substituting these into equation [17], the expected SSD simplifies to:

$$\mathbb{E}[SSD_S] = \frac{1}{n} \cdot [0.0543 \cdot f^2 \cdot (n - 1) + 0.1087 \cdot f + 0.8345] \quad (18)$$

This closed-form expression can be inverted to compute the ECP by solving for the value of $f$ such that $\mathbb{E}[SSD_S] = SSD_D$, where $SSD_D$ is the observed sum of squared deviations for the empirical dataset.

*A.3. Computation of ECP from the MAD*

The Mean Absolute Deviation (MAD) is computed as the average of the absolute deviations from BL, that is:

$$MAD = \frac{1}{9}\sum_{d=1}^{9} AbsDif_d = \frac{1}{9}\sum_{d=1}^{9}|Dif_d| = \frac{1}{9}\sum_{d=1}^{9}|O_d - B_d| \quad (19)$$

where $Dif_d = O_d - B_d$, $AbsDif_d$ is the absolute value of $Dif_d$, and $|\cdot|$ denotes the absolute value operator.

Since $O_d$ is approximately normally distributed, $Dif_d$ also follows a normal distribution with mean and variance:

$$\mathbb{E}[Dif_d] = p_d - B_d = f \cdot \delta_d \quad (20)$$

$$\text{Var}[Dif_d] = \frac{p_d \cdot (1 - p_d)}{n} \quad (21)$$

As $AbsDif_d$ is defined as the absolute value of a normally distributed variable, it follows a folded normal distribution, whose expected value is given by:

$$\mathbb{E}[AbsDif_d] = \sqrt{\frac{p_d \cdot (1-p_d)}{n}} \cdot \sqrt{\frac{2}{\pi}} \cdot exp\left(-\frac{(f \cdot \delta_d)^2}{2 \cdot \frac{p_d \cdot (1-p_d)}{n}}\right) + (f \cdot \delta_d) \cdot \left[2\Phi\left(\frac{f \cdot \delta_d}{\sqrt{\frac{\pi_d \cdot (1-\pi_d)}{n}}}\right) - 1\right]$$

(22)

where $\Phi$ represents the cumulative distribution function of the standard normal distribution. The expected value of the MAD for sample $S$ is then:

$$\mathbb{E}[MAD] = \frac{1}{9}\sum_{d=1}^{9} E[AbsDif_d]. \tag{23}$$

Expressions (22) and (23) capture how the expected MAD evolves as a function of the sample size $n$ and the contamination proportion $f$. Although this relation is more complex than in the case of the chi-squared or SSD statistics, it still allows for the computation of the ECP by determining the value of $f$ that satisfies $\mathbb{E}[MAD_S] = MAD_D$, where $MAD_D$ is the MAD observed in the empirical dataset. In this case, the ECP cannot be derived analytically but can be estimated numerically.

For consistency with the previous sections, it would be natural to present the expression of $\mathbb{E}[MAD_S]$ when the uniform distribution is used as the contaminating distribution. However, since this case does not lead to a simplified analytical form—unlike the chi-squared and SSD statistics—the general expression provided above remains the most informative and practical for computing the ECP. The required values of $B_d$, $\delta_d$, and $p_d$ can be easily computed for each digit, and the resulting expression can be efficiently evaluated numerically.

### A.4. Computation of ECP from the ED

The Euclidean Distance (ED) can be computed as:

$$ED = \sqrt{\sum_{d=1}^{9} E(O_d - B_d)^2}. \tag{24}$$

Unlike the previous divergence measures, ED is not an additive function, as it involves a square root operation over the sum of the squared differences. Consequently, it does not admit a closed-form expression for its expectation. However, since ED is a smooth, positive, and strictly increasing function of SSD, its expectation can be approximated by applying a first-order Taylor expansion around the expectation of SSD:

$$\mathbb{E}[ED] \approx \sqrt{\mathbb{E}[SSD]} \qquad (25)$$

As with the other measures, this approximation can be used to estimate the contamination proportion *f* that makes the expected ED for the hypothetical sample *S* equal to the observed ED in the empirical dataset *D*. However, it is important to note that this estimation does not rely on an exact analytical expression for the expectation and is therefore inherently less precise than the ECP derived from statistics like $\chi^2$, MAD, or *SSD*. For this reason, this estimation of ECP from ED should be interpreted with caution.

Expression (18) indicates the value of $\mathbb{E}[SSD]$ when the uniform distribution is chosen as the non-BL distribution. Substituting it into equation (25):

$$\mathbb{E}[ED] \approx \sqrt{\frac{1}{n} \cdot [0.0543 \cdot f^2 \cdot (n-1) + 0.1087 \cdot f + 0.8345]} \qquad (26)$$

This approximation enables a numerical estimation of the ECP for ED. While it may be sufficiently accurate for many practical applications, particularly if *n* and/or *f* are relatively high,[12] it does not yield the exact value of the expected ED. In cases

---

[12] To evaluate the accuracy of the approximation in expression (25), I performed Monte Carlo simulations with 10,000 replications for each combination of sample size $n \in \{100, 500,$

where both *n* and *f* are low, it may be preferable to use the simulation-based method described in section 2.2.5.

*A.5. Computation of ECP from other non-additive statistics*

In addition to the divergence measures reviewed in the previous subsections, researchers have assessed conformity with BL using other tests statistics, such as the Kolmogorov-Smirnov (KS), Kuiper, and Cramér-von Mises (CvM) statistics. Since these measures are non-additive, deriving a closed-form expression for $\mathbb{E}[T(S)]$ is considerably more complex than for the statistics discussed earlier. For this reason, no analytical expression has been derived here. Nevertheless, I propose an alternative numerical method to estimate the ECP from these non-additive statistics.[13]

Let *T(D)* denote the value of the divergence statistic observed in *D*. The procedure starts by evaluating the statistic *T* under the two extreme cases: *f*=0 (yielding $T_{f=0}$) and *f*=1 (yielding $T_{f=1}$). If *T(D)* is lower (higher) than $T_{f=0}$ ($T_{f=1}$), the ECP is set to 0 (1) accordingly. When *T(D)* lies within the range defined by $T_{f=0}$ and $T_{f=1}$, the procedure operates as follows:

---

1000, 5000, 10000} and contamination proportion *f* ϵ {0%, 5%, 10%, 15%, …, 100%}. In each simulation, a sample of *n* first-digit values was generated such that a proportion *f* followed a uniform distribution (used as the contamination distribution), and the remaining 1 – *f* observations followed BL. The ED was computed for each sample, and the approximation error was assessed by comparing the simulated mean of ED with the value predicted by expression (25). The results show that the relative error decreases as either the sample size *n* or the contamination proportion *f* increases.

[13] The supplementary material provides a fully documented Python script (ECP.py) that implements this estimation method, applied to the Euclidean Distance, Kolmogorov-Smirnov, Kuiper, and Cramér-von Mises statistics, as well as the estimation procedures for the other measures based on their closed-form expectations.

(1) Initialize a value for the contamination proportion *f*.

(2) Simulate a large number of samples of size *n*, each with contamination proportion *f*. Compute the average value of the divergence statistic across the simulated samples.

(3) Compare this average to *T(D)*. If the difference is not statistically significant, based on a test of means at a predefined significance level, the procedure terminates and the current value of *f* is returned as the estimated ECP.

(4) Otherwise, update the value of *f* to bring the expected statistic closer to *T(D)* and return to step 2.

Although computationally more intensive than closed-form derivations, this approach extends the applicability of the ECP framework to any divergence statistic that varies continuously and monotonically with the contamination proportion *f*, particularly in cases where no closed-form expression for $\mathbb{E}[T(S)]$ is available.